\documentclass[aps,pra,superscriptaddress,twocolumn,longbibliography, 10pt]{revtex4-1}
\usepackage[bookmarks=true,colorlinks,citecolor=blue,urlcolor=blue]{hyperref}
\usepackage{ucs}
\usepackage[english]{babel}
\usepackage[usenames, dvipsnames]{xcolor}
\usepackage[export]{adjustbox}
\usepackage{amsmath, 
			amsfonts,
			booktabs,
            microtype,
            mathtools, 
            graphicx,
            color,
            soul,
            nicefrac,
            float,
            url, 
            placeins,	
			braket,
			appendix} 	

\newcommand{\mb}[1]{{\color{teal} #1}}

\DeclareRobustCommand{\mb}[1]{%
  \ifmmode\text{\hl{$#1$}}\else\hl{#1}\fi
}

\begin{document}
\title{Preparation of hundreds of microscopic atomic ensembles in optical tweezer arrays}

\author{Y. Wang}\affiliation{ISIS (UMR 7006) and IPCMS (UMR 7504), University of Strasbourg and CNRS, 67000 Strasbourg, France}
\author{S. Shevate}\affiliation{ISIS (UMR 7006) and IPCMS (UMR 7504), University of Strasbourg and CNRS, 67000 Strasbourg, France}
\author{T. M. Wintermantel}\affiliation{ISIS (UMR 7006) and IPCMS (UMR 7504), University of Strasbourg and CNRS, 67000 Strasbourg, France}\affiliation{Physikalisches Institut, Universit\"at Heidelberg, Im Neuenheimer Feld 226, 69120 Heidelberg, Germany}
\author{M. Morgado}\affiliation{ISIS (UMR 7006) and IPCMS (UMR 7504), University of Strasbourg and CNRS, 67000 Strasbourg, France}
\author{G. Lochead}\affiliation{ISIS (UMR 7006) and IPCMS (UMR 7504), University of Strasbourg and CNRS, 67000 Strasbourg, France}
\author{S. Whitlock}\affiliation{ISIS (UMR 7006) and IPCMS (UMR 7504), University of Strasbourg and CNRS, 67000 Strasbourg, France}
\date{\today}

\begin{abstract}
We present programmable two-dimensional arrays of microscopic atomic ensembles consisting of more than 400 sites with nearly uniform filling and small atom number fluctuations. Our approach involves direct projection of light patterns from a digital micromirror device with high spatial resolution onto an optical pancake trap acting as a reservoir. This makes it possible to load large arrays of tweezers in a single step with high occupation numbers and low power requirements per tweezer. Each atomic ensemble is confined to $\sim 1\,\mu$m$^3$ with a controllable occupation from 20 to 200 atoms and with (sub)-Poissonian atom number fluctuations. Thus they are ideally suited for quantum simulation and for realizing large arrays of collectively encoded Rydberg-atom qubits for quantum information processing.
\end{abstract}

\maketitle

Neutral atoms in optical tweezer arrays have emerged as one of the most versatile platforms for quantum many-body physics, quantum simulation and quantum computation~\cite{Dumke2002,Bergamini_04,Nogrette_2014,Barredo_2016,Endres1024,Kim2016,Tamura2016,Norcia2018,Cooper2018,Anderegg1156}. This is largely due to their long coherence times combined with flexible configurations and controllable long-range interactions, in particular using highly excited Rydberg states~\cite{Saffman_2016,labuhn2016,Lienhard2018,Sanchez2018,Kim2018,Graham2019,deLeseleuc2019}. To date, substantial experimental efforts have been devoted to create fully occupied atomic arrays with $\simeq$1 atom in each tweezer by exploiting light-assisted inelastic collisions~\cite{Nicolas_2001,Grunzweig2010,Lester_2015} and rearrangement to fill empty sites~\cite{Barredo_2016,Endres1024,Barredo_2018,kumar2018sorting,Brown_2019,Ohl2019}. In combination with high-fidelity Rydberg blockade gates~\cite{Lukin2001,Isenhower2010,Wilk2010,Maller2015,Zeng2017,Levine2018,Levine2019,Graham2019}, these systems have been recently used to demonstrate coherent quantum dynamics of up to 51 qubits~\cite{bernien2017} and entangled states of up to 20 qubits in one-dimensional (1D) chains~\cite{Omran570}. So far however, achievable array sizes are limited to $\lesssim 100$ fully occupied sites (including for 2D and 3D systems), in part due to high power requirements and increasing complexity associated with the rearrangement process for larger arrays~\cite{Barredo_2018,Brown_2019,Ohl2019}.

\begin{figure}[t!]
	\includegraphics[width=0.46\textwidth]{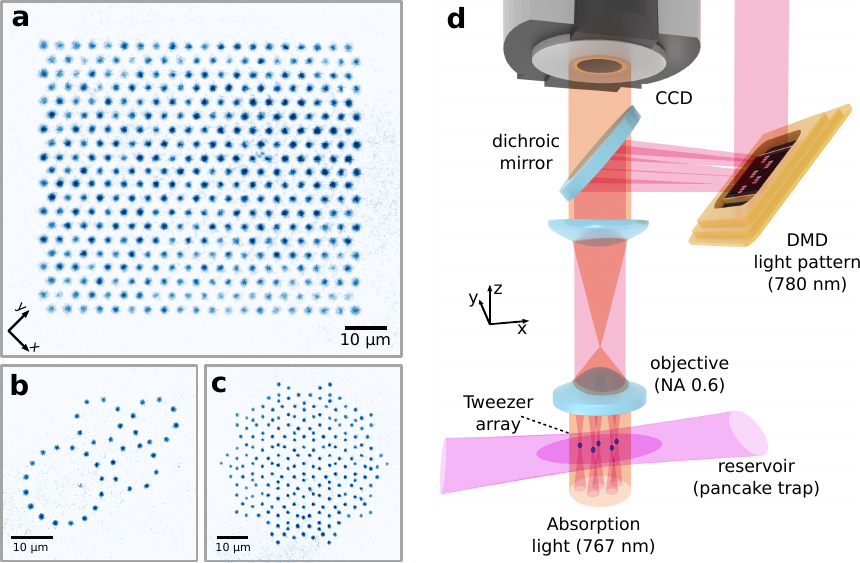}
	\caption{\textbf{Realization of a large tweezer arrays with large occupation number in each tweezer.} (a) Experimental absorption image of a 400 site triangular lattice, where each dark spot corresponds to a microscopic ensemble of $\approx 30$ ultracold $^{39}$K atoms. The lattice spacing is $4\,\mu$m and the apparent size of each spot is $\sim 0.75\,\mu$m ($e^{-1/2}$ radius), mostly limited by recoil blurring during imaging. (b) 40 site ring structure and (c) 226 site Penrose quasicrystal lattice. To improve the signal-to-noise ratio each image is an average of 20 absorption images. (d) Setup used to produce and load the tweezer arrays by projecting light from a digital micromirror device directly onto the atoms confined in an optical reservoir trap.}
	\label{fig:DMD_setup}
\end{figure}

In this paper we demonstrate an alternative approach to prepare large and uniformly filled arrays of hundreds of tweezers with large occupation numbers in a single step. This is exemplified by the 400 site triangular array shown in Fig.~\ref{fig:DMD_setup}a, as well as more exotic geometries such as connected rings (Fig.~\ref{fig:DMD_setup}b) and quasi-ordered geometries (Fig.~\ref{fig:DMD_setup}c) which exhibit structures on different length scales making them difficult to produce using other methods. Our approach involves transferring ultracold atoms from a quasi-2D optical reservoir trap into an array of optical tweezers produced by a digital micromirror device (DMD). To realize large arrays we optimize the loading process and the homogeneity across the lattice by adapting the DMD light patterns to control the trap depth of each tweezer. Each atomic ensemble is localized well within the typical Rydberg blockade radius and the typical interatomic separations of several micrometers are compatible with Rydberg-blockade gates. Furthermore, we show that the fluctuations of the number of atoms in each tweezer is comparable to or below the shot-noise limit for uncorrelated atoms. This makes the system well suited for quantum simulation of quantum spin models~\cite{Glaetzle2014,vanBijnen2015,Whitlock_2017,Kiffner2017,Letscher_2017,Zeiher2017,Lienhard2018,Sanchez2018,Leseleuc2019} and dynamics~\cite{Gunter2013,Robicheaux2014,Barredo_2015,Schempp2015,Schonleber2015,plodzien2018,Whitlock2019,Yang2019} in novel geometries, as well as for realizing quantum registers with collectively enhanced atom-light interactions for quantum information processing~\cite{Beterov2013,Ebert2015,Saffman_2016,Brion2007,Whitlock_2017, Wintermantel2019}. 

Our experimental cycle starts with a three-dimensional magneto-optical trap (MOT) loaded from a beam of $^{39}$K atoms produced by a two-dimensional MOT. This is overlapped with a far off-resonant pancake-shaped reservoir trap created by a 1064\,nm single mode laser with a power of $16\,$W tightly focused by a cylindrical lens. The beam waists are $\omega_z = 7.6\,\mu$m, $\omega_x = 540\,\mu$m and $\omega_y = 190\,\mu$m and the estimated trap depth is $330\,\mu$K. To maximize the number of atoms in the reservoir we apply an $8\,$ms gray-molasses cooling stage on the $\text{D}_1$ transition~\cite{Salomon2013}, yielding $3.3\times 10^5$ atoms in the $|4s_{1/2},F=1\rangle$ state at an initial temperature of $45\,\mu$K. 

Next we transfer the atoms to the tweezers from the reservoir trap. To generate the tweezers we illuminate a DMD with a collimated 780\,nm Gaussian light beam with a 4.3\,mm waist and a peak intensity of 1.44\,W/cm$^{2}$. We directly image the DMD plane onto the atoms (similar optical setup to Refs.~\cite{Muldoon_2012,Gauthier16}) with a calibrated demagnification factor of 53, using a 4f optical setup involving a 1500\,mm focal length lens and a 32\,mm focal length lens (Fig.~\ref{fig:DMD_setup}d). The latter is a molded aspheric lens with a numerical aperture of 0.6, located inside the vacuum chamber. With this setup, each $(13\,\mu$m$)^2$ pixel of the DMD corresponds to $(245\,$nm$)^2$ in the atom plane.

The DMD can be programmed with arbitrary binary patterns to control the illumination in the atom plane. To generate the tweezer arrays shown in Fig.~\ref{fig:DMD_setup}a-c we create different patterns of spots where each spot is formed by a small disk-shaped cluster of typically $A=20-100$ pixels. To detect the atoms we use the saturated absorption imaging technique~\cite{Reinaudi_2007,Ockeloen2010}. The probe laser is resonant to the $4s_{1/2}\rightarrow 4p_{3/2}$ transition of $^{39}$K at 767\,nm and with an intensity of $I\approx 2.1I_\text{sat}^\text{eff}$.
The atoms are exposed for $10\,\mu$s and the absorption shadow is imaged onto a charge-coupled-device camera using the same optics as for the DMD light patterns~(Fig.~\ref{fig:DMD_setup}d). The resulting optical depth image is well described by a sum of two-dimensional Gaussian distributions. By fitting and integrating each distribution we determine the number of atoms in each tweezer. By analyzing a background region of the images we infer a single-shot detection sensitivity of $3.9(5)$~atoms.

\begin{figure}[t!]
  \centering
	\includegraphics[width=1\columnwidth]{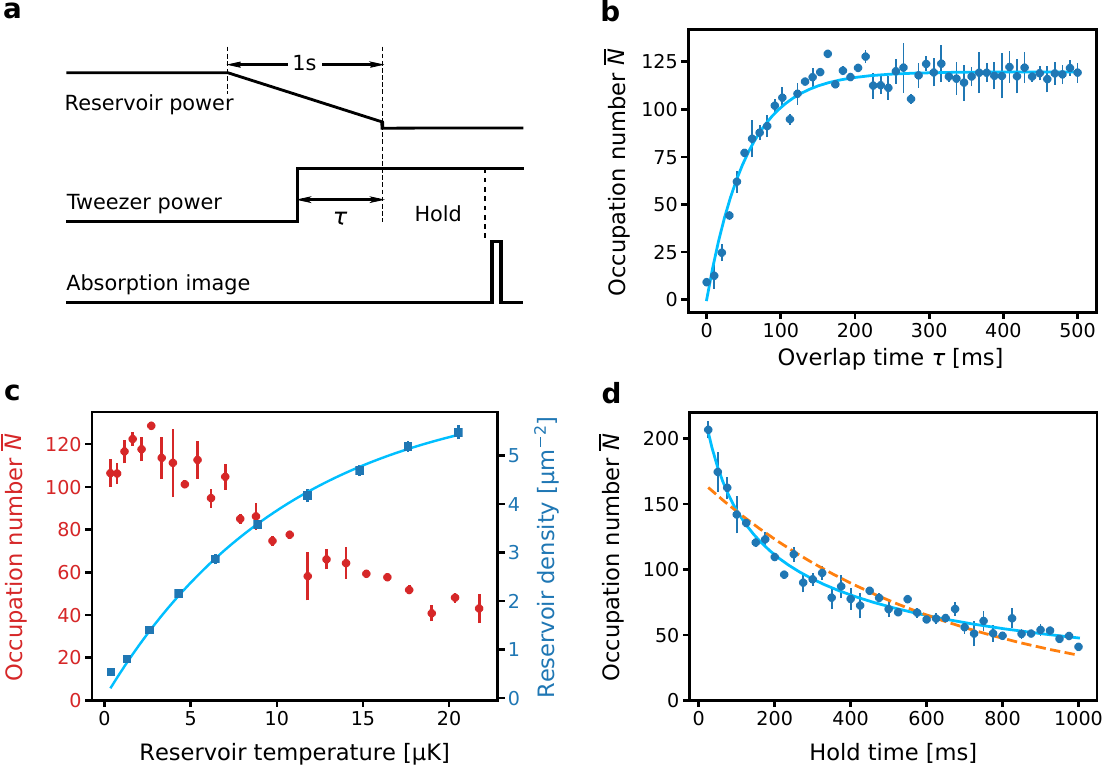}
	\caption{\textbf{Characterization of the loading process and lifetime of atoms in a single tweezer.} (a) Sketch of the experimental sequence used to load the tweezers. (b) Mean occupation number in a single tweezer as a function of the overlap time $\tau$ between the end of the reservoir evaporation ramp and turning on the tweezer. For $\tau\approx 200$\,ms the occupation number reaches its maximum value of $\bar N = 120$. (c) Mean occupation and two-dimensional density of atoms in the reservoir as a function of the reservoir temperature after evaporation. Optimal loading is achieved for a final reservoir temperature of $2\,\mu$K, which is a compromise between temperature and the remaining density of atoms in the reservoir. (d) Measurement of the lifetime of atoms held in the tweezer. The solid line is a fit to a model accounting for one- and three-body loss processes, while the dashed line is an exponential fit assuming one-body loss only. In (b),(c) and (d) the error bars depict the standard deviation over three experimental repetitions.}
	\label{fig:Loading}
\end{figure}

We found that optimal loading of the tweezers is achieved by evaporatively cooling the atoms in the reservoir trap while superimposing the DMD light pattern (Fig.~\ref{fig:Loading}a). At the end of the evaporation ramp the reservoir trap can be switched off leaving the atoms confined by the tweezers alone. The overall cycle time including MOT loading, evaporative cooling, transfer to the tweezers and imaging is $<4\,$s. Figures~\ref{fig:Loading}b,c show the characterization of the loading process for a single tweezer with $A=100$ pixels, corresponding to an optical power of $90\,\mu$W. 
There is little difference if the tweezer is switched on suddenly or ramped slowly, however we found it is beneficial to turn on the tweezers at least $200\,$ms before the end of the evaporation ramp (Fig.~\ref{fig:Loading}b), indicating a significant enhancement of the loading through elastic collisions with the reservoir atoms~\cite{Comparat2006}. Figure \ref{fig:Loading}c shows that the mean occupation number $\bar N = \langle N\rangle_i$ (with $i$ denoting different experimental realizations) strongly depends on the final temperature of the reservoir, with the maximal $\bar N=120(5)$ found for $T_\text{res}\approx 2\,\mu$K. The temperature of the atoms after loading measured using the time-of-flight method for a single tweezer is $17(1)\,\mu$K. Figure~\ref{fig:Loading}d shows the lifetime of the atoms in the tweezer. An exponential fit describing pure one-body decay (dashed curve) is clearly ruled out by the data while a model which includes both three-body and one-body decay~\cite{Whitlock_2010} provides an excellent fit (solid curve). From this model we extract both the three-body and one-body decay constants $k_3^{-1} = 110\,\text{ms}$, $k_1^{-1} = 3100\,\text{ms}$, which are both orders of magnitude longer than typical timescales in Rydberg atom experiments.

\begin{figure*}[ht!]
  \centering
	\includegraphics[width=0.70\textwidth]{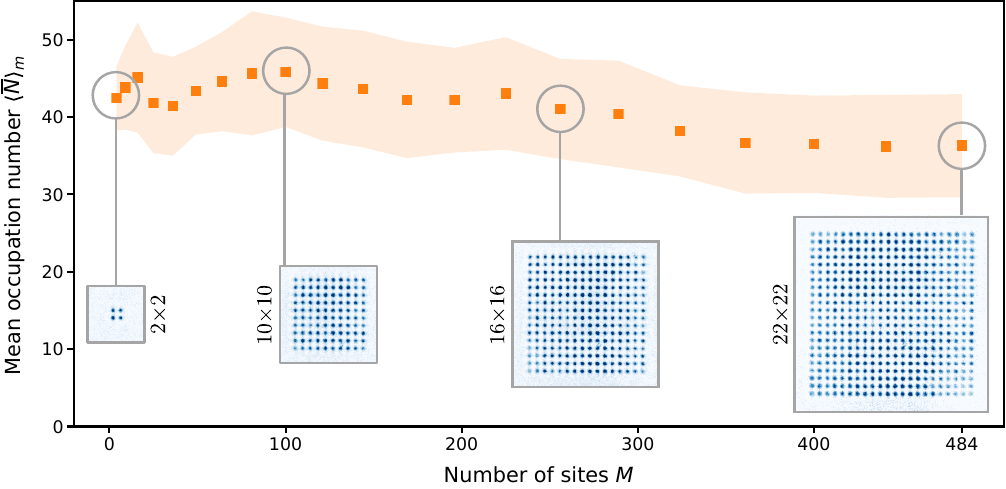}
	\caption{\textbf{Scaling up to hundreds of tweezers.} The square data points show the array-averaged mean occupation number for square arrays with a period of $3.5\,\mu$m and different numbers of sites ranging from $M=2\times 2$ to $M=22\times 22$. Over this range the array averaged occupation number $\langle \bar N\rangle_m \approx 40$ is mostly homogeneous and insensitive to the number of tweezers. The shaded region represents the uniformity of each array, computed as the standard deviation of the mean occupation number. The insets show exemplary absorption images for 4 different sized arrays, each averaged over 20 experimental repetitions.}
	\label{fig:examples}
\end{figure*}

We now show that it is straightforward to scale up to a large number of sites while maintaining a uniformly high occupation of each tweezer. For arrays with more than approximately $M=(10\times 10)$ sites we found it is beneficial to adapt the DMD pattern to compensate the Gaussian illumination profile. We adapt the number of pixels in each cluster according to its distance from the center of the illumination region following an inverted Gaussian dependence $A_m = A_\text{min}\exp(g r_m^2)$ where $m$ indexes a tweezer at position $r_m$ in the DMD plane. The parameters $A_\text{min}=20$ pixels and $g=2.3\times 10^{-5}$ were manually adapted to obtain approximately equal optical depths for each tweezer, where $A_\text{min}=20$ corresponds to an optical power per tweezer of $18\,\mu$W. The maximum array size of $M=(22\times 22)$ was fixed such that $\text{max}(A_m) = 66$ pixels, due to larger spots increasing the apparent size of the atomic ensembles near the edge of the array. Once the optimal compensation profile is found it can be applied to different geometries without further adjustment.

Figure~\ref{fig:examples} shows the array averaged mean occupation number $\langle \bar N\rangle_m$ for square arrays with different numbers of sites. The solid orange symbols show the occupation number averaged over the entire lattice and over 20 experimental repetitions. The experiments show that $\langle\bar N\rangle_m \simeq 40$ is approximately constant for tweezer arrays with different numbers of sites $4\leq M \leq 484$. As an example, for $M=400$ the standard deviation calculated from the average of $20$ images is $0.17\langle \bar N\rangle_m$, compared to $0.55\langle \bar N\rangle_m$ without any compensation. The uniformity could be further improved by adapting $A_m$ for each tweezer individually, but it is already better than the expected intrinsic shot-to-shot fluctuations of the atom number in each site due to atom shot-noise for $\bar N>36$.

The apparent size of each atomic ensemble, found by analyzing the averaged absorption images, ranges from $0.64\,\mu$m near the center of the field to $0.93\,\mu$m at the edges ($e^{-1/2}$ radii of each absorption spot), limited by recoil blurring, the finite resolution of the imaging system including off axis blurring and the finite size of each DMD spot. By projecting the DMD light pattern onto a camera in an equivalent test setup we independently determine the beam waist of each tweezer to be $0.9\,\mu$m. Assuming each tweezer is described by a Gaussian beamlet and approximating the atomic cloud by a thermal gas with a temperature $\sim V_0/5$ (with trap depth $V_0$), we infer a cloud size of $\sigma_{r,z} = \{0.2,1.0\}\,\mu$m. This is reasonably close to an independent estimate $\sqrt[3]{\sigma_r^2\sigma_z}\approx 0.6\,\mu$m based on the experimentally measured three-body loss rate and a theoretical calculation of the zero field three-body loss coefficient for $^{39}$K~\cite{Esry_1999}. 

The small spatial extent of each ensemble is encouraging for experiments which aim to prepare a single Rydberg excitation at each site, as it is significantly smaller than the nearest neighbour distance and the typical Rydberg blockade radius of $R_{bl}\sim 3-6\,\mu$m (depending on principal quantum number). Approximating the density distribution as quasi-one-dimensional we estimate the fraction of blockaded atoms $f_{bl}=\text{erf}(R_{bl}/2\sigma_z)$ where $\text{erf}(x)$ is the Gauss-error function. For $R_{bl}/\sigma_z\geq 3$, $f_{bl}\geq 0.97$ which suggests that the blockade condition within a single tweezer should be well satisfied. To serve as effective two-level systems (comprised of the collective ground state and the state with a single Rydberg excitation shared amongst all atoms in the ensemble), it is additionally important that the fluctuations of the atom number from shot-to-shot are relatively small otherwise the $\sqrt{N}$ collective coupling~\cite{Dudin2012,Ebert2015, Zeiher2015} will reduce the single atom excitation fidelity. Previous theoretical estimates have assumed Poisson distributed atom shot noise~\cite{Whitlock_2017}, which we generalize to the case of a stretched Poissonian distribution and imperfect blockade. However we neglect other possible imperfections such as spectral broadening due to laser linewidth or interactions between ground state and Rydberg atoms~\cite{Whitlock_2017}. We assume that within the blockade volume, the probability to excite a single atom undergoes collective Rabi oscillations $p_i = 1-\text{cos}^2(\sqrt{N_i}\Omega t/2)$ where $\Omega$ is the single atom Rabi frequency~\cite{Whitlock_2017}. In contrast the non-blockaded fraction of atoms $1-f_{bl}$ undergoes Rabi oscillations at the frequency $\Omega$. By expanding around the time for a collective $\pi$ pulse: $t= \pi/(\sqrt{\bar N}\Omega)$ and averaging over a stretched Poissonian statistical distribution for atom number fluctuations we find that the infidelity for producing exactly one excitation is $\epsilon \approx (\pi^2/4)\left[\text{var}(N)/(4\bar N^2)+(1-f_{bl})\right]$. For $f_{bl}\rightarrow 1$ the infidelity is proportional to the relative variance $\text{var}(N)/\bar N^2$.

\begin{figure}[t!]
  \centering
	\includegraphics[width=0.9\columnwidth]{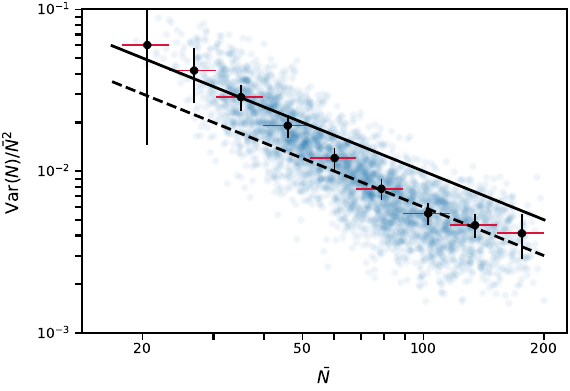}
	\caption{\textbf{Relative variance of the atom number in each tweezer $ \mathrm{var}(N)/\bar N ^2$ as a function of the mean occupation number $\bar N$.}  The blue points show the relative variance computed from 20 repetitions of the experiment and for different mean occupation numbers. The black symbols show the same data after binning, where the bin widths are indicated by the horizonal red bars. The vertical error-bars show the standard error computed over the values inside each bin multiplied by a scaling factor of 10 for better visibility. The solid black line is a prediction assuming Poissonian atom number fluctuations $ \text{var}(N)/\bar N ^2 = 1/\bar N$ and the dashed black line is a prediction assuming sub-Poissonian atom number fluctuations $ \text{var}(N)/\bar N ^2 = 0.6/\bar N$.}
	\label{fig:fluc}
\end{figure}

To estimate the relative variance of the atom number fluctuations we prepare tweezer arrays with different numbers of sites and trap depths, corresponding to mean occupation numbers from $\bar N = 20$ to $\bar N = 200$. For each set of experimental conditions we take 20 absorption images from which we compute the mean occupation and the variance of the atom number in each tweezer. Fig.~\ref{fig:fluc} shows the relative variance $\text{var}(N)/ \bar N ^2$ calculated for 75880 tweezer realizations. We see that for smaller atom numbers the relative variance is consistent with the expected Poissonian atom shot noise for independent particles [$\text{var}(N) = \bar N$, shown by the solid black line], while for $\bar N\gtrsim 50$ the fluctuations are sub-Poissonian, reaching the lower limit expected for three-body loss [$\text{var}(N) = 0.6\bar N$, shown by the solid dashed line]~\cite{Whitlock_2010}. For $\bar N>40$ the expected infidelity due to atom number fluctuations would be below 0.03, showing that this system should be compatible with high fidelity preparation of individual Rydberg excitations and quantum logic gates, and could still be further improved using adiabatic or composite pulse techniques~\cite{Beterov2013}.

In conclusion, we have demonstrated an approach for realizing hundreds of ultracold atomic ensembles in programmable two-dimensional arrays, where each tweezer has approximately uniform filling, small spatial extent and small fluctuations of the atom number from realization to realization. Compared to stochastic loading of the tweezers via light assisted collisions from a MOT this has several advantages. First, it is possible to achieve very high occupation numbers $ N\gg 1$ with relatively low power requirements per tweezer, since the temperature of the initial reservoir trap can be lower than the typical temperatures in a MOT and elastic, rather than inelastic, collisions lead to a high filling probability. This is beneficial for scaling up to hundreds or even thousands of tweezers as the large volume of the reservoir trap makes it possible to simultaneously fill many tweezers in parallel without the need for additional lasers and complex rearrangement protocols to fill empty sites. Additionally, atomic ensembles present the possibility to evaporatively cool the atoms in each tweezer to reach high phase space densities or as a more controlled starting point for (quasi)deterministic single atom preparation schemes using controlled inelastic collisions~\cite{Grunzweig2010,Brown_2019} or the Rydberg blockade effect~\cite{Ebert2014,Weber2015,Whitlock_2017}. The observation that the number of atoms inside each tweezer exhibits fluctuations below the Poissonian atom shot noise limit is especially promising for quantum information processing based on small Rydberg blockaded atomic ensembles benefiting from fast collectively enhanced light-matter couplings~\cite{Ebert2014, Brion2007, Saffman_2016,Whitlock_2017,Wintermantel2019}.

\acknowledgements{This work is supported by the `Investissements d'Avenir' programme through the Excellence Initiative of the University of Strasbourg (IdEx), the University of Strasbourg Institute for Advanced Study (USIAS) and is part of and supported by the DFG Collaborative Research Center `SFB 1225 (ISOQUANT)'. S.S., T.M.W. and M.M. acknowledge the French National Research Agency (ANR) through the Programme d'Investissement d'Avenir under contract ANR-17-EURE-0024.}

\bibliography{Microtrap.bib}

\end{document}